\documentclass[a4paper,12pt]{article}
\usepackage{amsmath,amsfonts,amssymb,cite}

\begin{document}

\begin{center}
{\Large\bf Regge trajectories in light and heavy mesons: the pattern of appearances and possible dynamical explanations\footnote{Presented at HADRON 2019.}}
\end{center}

\begin{center}
{\large S. S. Afonin}
\end{center}

\begin{center}
{\it Saint Petersburg State University, 7/9 Universitetskaya nab.,
St.Petersburg, 199034, Russia\\
E-mail: s.afonin@spbu.ru}
\end{center}

\begin{abstract}
I briefly review the Regge approach to the hadron spectrum and advocate a dynamical emergence of principal quantum number in the known spectrum of light non-strange mesons. Further it is shown how the linear radial trajectories with universal slope can be extended to heavy quarkonia and a qualitative string interpretation is given. After that I discuss a recently proposed non-string model leading to a natural appearance of linear Regge trajectories and explaining many mass relations.
\end{abstract}

\section{Regge trajectories and degeneracy in light meson spectrum}

The Regge phenomenology emerged from the dual
amplitudes~\cite{collins} and gave rise to various hadron string
models. This approach continues to play an important role in the
study of hadron spectroscopy. The dual amplitudes and some related
string approaches predict the following behavior of meson masses,
\begin{equation}
\label{1}
M_n^2=a(J+bn+c), \qquad J,n=0,1,2,\dots,
\end{equation}
where $J$ is the spin, $n$ enumerates the radially excited states,
$a$ represents a universal slope and $b,c$ are some constants. The
relation~\eqref{1} reproduces the classical Regge behavior
$M^2\sim J$ observed in the light baryons and mesons. Also this
relation predicts the equidistant daughter Regge trajectories. The
number $n$ enumerates the states on these "radial" Regge
trajectories. The available experimental data on light non-strange
mesons seem to confirm the linear Regge
behavior~\eqref{1}~\cite{ani,bugg,klempt,Li,cl1,cl1b,cl2,cl2b,cl2c,cl2d,cl2e,cl2f,cl2g}.

A popular assumption about the
structure of mesons is the hypothesis that the mesons represent a
gluon string with a quark/antiquark at the ends.
A careful analysis of light non-strange meson spectrum showed, however, that
instead of spin $J$ in Eq.~\eqref{1} one should use the angular momentum $L$
of the quark-antiquark pair~\cite{cl2,cl2b,cl2c},
\begin{equation}
\label{1b}
M^2\sim a(L+bn).
\end{equation}
In addition, the data suggest $b=1$ that is a remarkable property
leading to numerous Hydrogen-like degeneracies~\cite{cl3,cl3b,cl3c,cl3d}.

\section{Universal description}

Various attempts to generalize the relation~\eqref{1} to the
sector of heavy quarks resulted in the emergence of strong
non-linearities with respect to $J$ and $n$. Let us look, however,
at the experimental data. A relatively rich set of data on radial
recurrences in this sector exists only for the unflavored vector
heavy quarkonia. Using the relevant data from the Particle
Data~\cite{pdg}, see Refs.~\cite{AP,AP2}, one can see again the emergence
of linear behavior as a function of consecutive number $n$.
Aside from the ground state, the masses approximately lie on the
linear radial trajectories
\begin{equation}
\label{2}
M_n^2=a(n+b),
\end{equation}
where we re-denoted the vector intercept $b=1+c$. In general, the slope and
intercept in~\eqref{2} depend strongly on the quark flavor~\cite{AP,AP2}.

The universal linearity suggests that some
universal gluodynamics lies behind the observed behavior. We can
propose a simple string scheme explaining this universality. The
central idea is that at the conditions when a quark-antiquark pair
form a resonance in the QCD vacuum, the pair may be viewed as a
virtally static system. The binding is provided by the exchange of
some particle (one might speculate that this particle is a pomeron). 
And one must quantize the motion of this particle
(not the radial motion of quarks as in the standard hadron string
approaches!). The total energy (mass) of the system is
\begin{equation}
\label{3}
M=m_1+m_2+p+\sigma r.
\end{equation}
Here $m_1$ and $m_2$ are the masses of quark and antiquark
separated by the distance $r$, $p$ is the momentum of exchanged
particle, and $\sigma$ represents the standard string tension.
Let us apply the semiclassical quantization to the momentum $p$
\begin{equation}
\label{4}
\int_0^l p\,dr=\pi(n+b),\qquad n=0,1,2,\dots.
\end{equation}
Here $l$ is the maximal quark separation and the constant $b$
depends on the boundary conditions. Substituting $p$
from~\eqref{3} to~\eqref{4} and making use of the definition
$\sigma=\frac{M}{l}$ we obtain the linear radial trajectory
\begin{equation}
\label{5}
(M_n-m_1-m_2)^2=2\pi\sigma(n+b).
\end{equation}

In our unflavored case $m_1=m_2\equiv m$ and the relation~\eqref{5}
can be simplified to
\begin{equation}
\label{6}
(M_n-2m)^2=a(n+b),
\end{equation}
which is our generalization of the linear spectrum~\eqref{2} to
non-zero quark masses. In this relation, the universal
gluodynamics (the slope $a$) and dependence on quantum numbers
(the dimensionless intercept $b$) are clearly separated from the
contribution of quark masses. For this reason, the parameters $a$
and $b$ in the relation~\eqref{6} should be flavor-independent.
The analysis of Refs.~\cite{AP,AP2} indeed confirmed this expectation.
A successful check of relation~\eqref{5} in the case heavy-light systems
was performed in Ref.~\cite{Chen}.

\section{A possible non-string mechanism for linear Regge trajectories}

Usually the observation of linear
trajectories is interpreted as an evidence for string picture
of mesons. In spite of many attempts,
however,
a satisfactory quantized hadron string has not been constructed.
Among typical flaws of this approach one can mention the absence of
spontaneous chiral symmetry breaking, rapidly growing (with mass)
size of meson excitations, and totally
unclear role of higher Fock components in the hadron wavefunction.

A novel realization of quark model concept has been recently proposed in Ref.~\cite{mult}.
This realization leads to a natural
(and alternative to hadron strings) explanation of Regge recurrences
and is potentially free of typical shortcomings inherent to string,
potential and some other approaches.
Stated more strictly, the given approach represents a phenomenological mass counting
scheme in which the relativistic invariance, renormalization invariance,
chiral symmetry breaking, higher Fock components, and linear Regge
behavior are qualitatively built-in.
The approach allows to classify
the light mesons without use of any angular momentum associated
with hadron constituents. The states $\pi_1$
(which are exotic for the standard quark model) emerge in a natural
way while the other exotic quantum numbers remain forbidden. The
constructed mass counting scheme permits to obtain hadron masses
from very simple relations with a typical accuracy comparable to
numerical calculations in complicated dynamical models.

The underlying motivation is as follows. The hadron masses must be renorminvariant.
On the other hand, the light non-strange mesons can be viewed, in one way or another, as
some quantum excitations of pion. Basing on the Gell-Mann--Oakes--Renner relation for
the pion mass and the known formula for the stress-energy tensor in QCD it is motivated
that the masses of these mesons can be represented as
\begin{equation}
\label{2b}
m_h^2=\Lambda (E_h+2m_q)=\Lambda E_h+m_\pi^2.
\end{equation}
The energy parameter $E_h$ seems to be related to the renorminvariant gluon condensate, $E_h\sim\alpha_s\langle
G_{\mu\nu}^2\rangle/\langle\bar{q}q\rangle$~\cite{mult}. A phenomenological mass counting scheme based on the relation~\eqref{2b}
and a demonstration how it describes the meson spectroscopy is scrutinized in Ref.~\cite{mult}.

It is also suggested that the parameter $E_h$ can be interpreted as an
effective energy of some constituents different from the current quarks.
The problem is to propose a model for these constituents which should
represent some excitations inside the pion. Three basic
excitations are postulated. The first one appears when one of quarks absorbs a gluon
of certain energy $E_h=E_\rho$. The spin of excited quark changes its
direction to the opposite one, converting the original spin-singlet
$q\bar{q}$-pair ($\pi$-meson) to the spin-triplet one ($\rho$-meson).
The second kind of excitation emerges due to formation of
spin-singlet $q\bar{q}$-pair with effective mass $E_h=E_0$ (the lower
index stays for total spin of $S$-wave $q\bar{q}$-pair).
The third basic
excitation is the formation of
spin-triplet $q\bar{q}$-pair with effective mass $E_h=E_1$.
The formation of these constituent pairs are likely a QCD
analogue of formation of para- and ortho-positronia from photons.
By assumption, any excitation inside pion leading to an observable
resonance can be represented as a combination of these
basic excitations so that the total effective energy $E_h$ in~\eqref{2b}
is just a sum (appropriate number of times) of $E_\rho$, $E_0$, and $E_1$.
Simultaneously this will dictate the quantum numbers of constructed resonance.
In a sense, the introduction of these constituent pairs reflects the excitations
of higher Fock components in the pion wavefunction.
These pairs in excited hadrons,
in some sense, bear a superficial resemblance to neutrons and protons
in atomic nuclei.
The given approach represents thus a new
realization of the quark model concept, a realization in which
highly excited states appear due to multiquark components in
hadron wavefunctions.

The inclusion of strange quarks into this approach is straightforward.
It would be interesting to extend the underlying ideas to light
baryons and to hadrons with heavier quarks.

\section*{Acknowledgments}

The present work was supported by the Saint-Petersburg State University
travel grant (Id: 40114404).

\end{document}